 \documentclass[onecolumn, 12pt, draftclsnofoot, comsoc]{IEEEtran}

\usepackage{gensymb}
\usepackage{cite}
\usepackage{amsmath,amssymb,amsfonts}
\usepackage{graphicx}
\usepackage{xcolor}
\usepackage{kotex}  

\usepackage{caption}
\usepackage{subcaption}

\usepackage{stfloats}
\usepackage{float}
\usepackage{wrapfig}
\usepackage{xcolor}
\usepackage{colortbl} 
\usepackage{color,soul}
\usepackage{url}
\usepackage{verbatim}
\usepackage{graphicx}
\usepackage{cite}

\usepackage{multirow}
\usepackage{array,boldline, makecell,booktabs}
\usepackage{longtable} 

    \usepackage[group-separator={,},group-minimum-digits=4]{siunitx}

\begin{document}
%
\title{RIS-Empowered  LEO Satellite Networks for 6G: Promising Usage Scenarios and Future Directions}
%
%
%

\author{\IEEEauthorblockN{Mesut Toka, ‪Byungju Lee‬, Jaehyup Seong, Aryan Kaushik, Juhwan Lee, \\Jungwoo Lee, Namyoon Lee, Wonjae Shin, and H. Vincent Poor}\thanks{M. Toka is with Ajou University, South Korea. B. Lee is with Incheon University, South Korea. A. Kaushik is with University of Sussex, United Kingdom. J. Lee and J. Lee are with Seoul National University, South Korea. J. Seong, N. Lee, and W. Shin are with Korea University, South Korea. V. Poor is with Princeton University, USA. (Corresponding author: Wonjae Shin)}}

%
%


\maketitle


\begin{abstract}

Low-Earth orbit (LEO) satellite systems have been deemed a promising key enabler for current 5G and the forthcoming 6G wireless networks.
Such LEO satellite constellations can provide worldwide three-dimensional coverage, high data rate, and scalability, thus enabling truly ubiquitous connectivity.
On the other hand, another promising technology, reconfigurable intelligent surfaces (RISs), has emerged with favorable features, such as flexible deployment, cost \& power efficiency, less transmission delay, noise-free nature, and in-band full-duplex structure. LEO satellite networks have many practical imperfections and limitations; however, exploiting RISs has been shown to be a potential solution to overcome these challenges. Particularly, RISs have the capability of enhancing link quality, reducing the Doppler shift effect, and mitigating inter-/intra beam interference.
In this article, we delve into the exploitation of RISs in LEO satellite networks. First, we present a holistic overview of LEO satellite communication and RIS technology, highlighting potential benefits and challenges. Second, we describe promising usage scenarios and applications in detail. Finally, we discuss potential future directions and challenges on RIS-empowered LEO networks, offering futuristic visions of the upcoming 6G era.

\end{abstract}


\IEEEpeerreviewmaketitle

\section{Introduction }
\label{sec:1}

The explosive growth in the number of various form-factor devices globally, with the attendant tremendous growth in demand for high data rate and service continuity, has led to the exploitation of non-terrestrial networks (NTNs) in the fifth-generation (5G) and upcoming sixth-generation (6G) wireless networks. Satellite networks are promising enablers that allow 6G communication systems to provide broadband coverage owing to their scalability, reliability, and ubiquitous coverage. Satellites are categorized according to their orbital altitudes: Geostationary orbit (GEO), medium-Earth orbit (MEO), and low-Earth orbit (LEO) at altitudes of $\num{35786}$ km, $\num{2000}-\num{35786}$ km, and $\num{200}-\num{2000}$ km, respectively. Among these, LEOs have received the most interest 
{due to their key features, such as low-latency, flexibility for mega-constellations, better frequency reuse, and higher link quality.} 

Nevertheless, LEO satellites face several challenges, such as severe Doppler effects,
limited power budgets, large \& moving cells, and short visibility times \cite{YeJia}. Previous works have considered the application of pre/post-compensation to mitigate the Doppler effect
and effective beamforming \& multiple access (MA) techniques for improved power usage and interference management. Recently, a promising technology called \textit{reconfigurable intelligent surface} (RIS) has emerged as a key enabler for fulfilling the requirements of the envisioned LEO-based 6G communication applications. RISs have several potential advantages, including flexible deployment everywhere, cost and power efficiency, less transmission delay, and in-band full-duplex capabilities ranging from sub-6 gigahertz (GHz) to terahertz (THz), making them compatible with existing and envisioned technologies \cite{WuQ, Tekbiyik}. Indeed, the integration of RIS technology with LEO satellites simultaneously solves two problems in that RIS can combat the aforementioned challenges in an energy-efficient manner and lower the cost of infrastructure on the ground. For instance, RISs can resolve link failure issues in dense urban scenarios, where millimeter waves (mmWave) and/or THz beams are vulnerable to blockage. Additionally, the use of RISs in free-space optics and computer vision-aided wireless communication technologies are potential solutions for the backhaul of LEO satellites and the identification of blockage locations. These considerations motivate us to focus on the RIS-empowered LEO satellite networks since the recent advances in the industry have prompted various schemes of LEO constellations not only for very small aperture terminal (VSAT) but also direct-to-cell (D2C) connectivity.

This article aims to shed light on promising applications and future directions for RIS-empowered 6G LEO satellite networks. Differing from existing studies, we delve deeper into the key roles of RISs in integrated sensing and communication (ISAC), the Doppler compensation, and artificial intelligence (AI)-enabled ISAC in LEO satellite networks. To this end, we first present background on LEO satellite communication and highlight its key advantages and challenges. Next, we provide motivations for incorporating  RISs into LEO satellite networks and investigate promising use cases for RIS-empowered LEO satellite networks. Finally, future directions and
challenges are discussed to envision an interplay with different enabling technologies for 6G.

\section{LEO Satellite Communication for 6G} 
\label{sec:2}

We discuss the key advantages of LEO satellites and their technical challenges in the following subsections.

\subsection{Main Advantages of LEO Systems}
\label{sec:2.2}

\subsubsection{Low-Latency and Link Quality Compared to GEO} Compared to GEO, LEO satellites revolve in orbit closer to the surface of the Earth and can guarantee a faster response time. 
%
Specifically, a GEO satellite's round-trip delay exceeds 250 ms, whereas that of a LEO satellite at an altitude of 600 km is less than 30 ms \cite{3GPP_811}.
%

\subsubsection{{Global Coverage using LEO Constellations}} 
Due to the low-altitude orbits, the size of satellite's footprints decreases, resulting in more limited coverage. 
Global coverage would require the deployment of a mega-constellation consisting of a large number of LEO satellites. 

\subsubsection{Frequency Reuse} Owing to the smaller beam footprint of LEO satellites, frequency reuse would be more effective than that of GEO satellites. 
In addition, through the frequency reuse of multi-beam antennas, high-throughput satellites can achieve several times higher throughput than the conventional single-beam structure. To achieve the potential benefits of frequency reuse and multi-beam techniques, the coexistence of GEO and LEO satellites should be considered to enable aggressive frequency reuse.
%

\begin{figure*}[t!]
    \begin{center}    
        \includegraphics[width=1\linewidth]{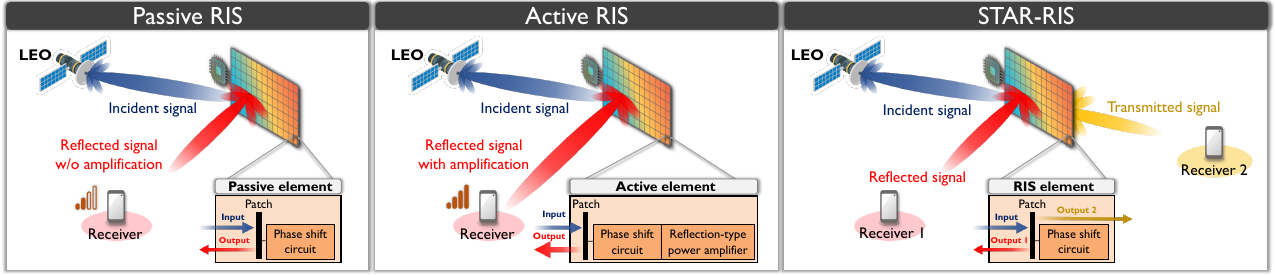}
    \end{center}
    \setlength{\belowcaptionskip}{-2pt}
    \caption{Comparison of passive RIS, active RIS, and STAR-RIS.}
    \label{fig1p}
\end{figure*}

\subsection{Key Challenges Facing 6G LEO Satellite Networks}
\label{sec:2.3}

\subsubsection{Doppler Effect} LEO satellites suffer from the Doppler effect because of high velocity relative to the ground. 
At an altitude of 600 km, the maximum Doppler shifts of a LEO satellite for 2 GHz and 20 GHz bands are 48 and 480 kHz, respectively, which are approximately 50 times bigger than that tolerated at 5G networks \cite{3GPP_811}.
Further, a differential Doppler shift can occur within the beam footprint of the LEO satellite according to the location of ground terminals. To resolve this issue, it is possible to perform the segment-wise pre-compensation for the Doppler shift at the beam center and the differential Doppler shift due to the difference between the user terminal’s position and the beam center.
However, this is still a challenging problem to be addressed. 

\subsubsection{Weak Line-of-Sight (LOS) Signal} One of the key features of satellite channels is that it is dominated by LOS components. 
The LOS probability depends on several factors, including the environment and the elevation angle (i.e., the angular height of the satellite measured from the receiver's horizontal line). 
For example, for an elevation angle of 10°, the LOS probability is lower in dense urban areas (28.2$\%$), whereas the LOS probability is higher in suburban and rural areas (78.2$\%$) \cite{3GPP_811}. Thus, a poor LOS probability brings several challenges to LEO satellite communication systems, such as poor link budget and severe signal attenuation. 

\subsubsection{Power Limit} 
LEO satellites consist of less mass to be cost-effective. Therefore, they have limited space to be equipped with solar panels and batteries, leading to a limited power budget \cite{YeJia}.
Specifically, the effective isotropically radiated power of the GEO satellite (e.g., 59 dBW/MHz) is much higher than that of the LEO satellite (e.g., 34 dBW/MHz). For efficient power consumption, transparent LEO satellites can be exploited to reflect the signal through the feeder link and forward it to the ground terminal. In contrast, to fully achieve low-latency applications, regenerative LEO satellites exploiting the inter-satellite links (ISL) can be exploited; however, they consume more power owing to on-board processing. Overall, effective power management of LEO satellites is critical to ensure their performance.

\subsubsection{Interference} 
LEO satellite systems should consider 3D network interference with the high-altitude of satellites. As the LEO satellite’s altitude increases, the beamwidth can be wider, thereby intensifying inter-beam interference. Considering a multi-layer structure where GEO and LEO satellites are combined, effective interference management can be accomplished by employing GEO satellites as nodes to control LEO satellites. 
To further manage interference caused by the coexistence of non-terrestrial and terrestrial networks, advanced MA techniques, such as space-division MA, non-orthogonal MA, and rate-splitting MA, can be applied.

\section{Motivations for Integrating RIS with LEO Satellite Networks 
} 

\label{sec:3}
The RIS technology has led to a rapid paradigm shift in 6G wireless networks since it can benefit use cases and enhancements of 5G, such as beam management, dynamic spectrum sharing, positioning, energy saving, NTNs, and so on, covered by 3GPP \cite{3GPPref}. To shed light on how RIS technology can be constructively exploited in LEO satellite networks, we discuss its basic concepts and potential benefits.



\subsection{Basic Concepts of RIS}

\label{sec:3a}

The RIS is an intelligent planar metasurface, which allows the propagation of electromagnetic waves to be controlled by inducing amplitudes and/or phase changes in the received signal. Each element of  RISs consists of micro-electronic devices or micro-electromechanical system switches, such as PIN diodes, varactors, and memristors that can be tuned in real-time to generate reflection coefficients \cite{WuQ}. 
RISs can be categorized as either \textit{active} or \textit{passive} in terms of their operating mode, as shown in Fig. \ref{fig1p}. 
Unlike passive RISs, active RISs have simple amplification circuits to overcome multiplicative fading effects that limit capacity gains \cite{ZhangActiveRIS}. 
In addition, two types of RIS technologies exist in terms of the reflection/refraction coverage: \textit{conventional} RIS, which only reflects electromagnetic waves within the half-space coverage, and \textit{simultaneous transmission and reflection} (STAR)-RIS \cite{LiuYSTAR} which not only reflects but also refracts electromagnetic waves within full-space $360^{\circ}$ coverage. Unlike \textit{conventional} RISs, STAR-RISs introduce both electric- and magnetic-currents, allowing some part of the incident signal to pass through the metasurface. Thus, the most beneficial feature is that both outdoor and indoor users can be served simultaneously.



\begin{figure*}[t!]
    \begin{center} 
        \includegraphics[width=1\linewidth]{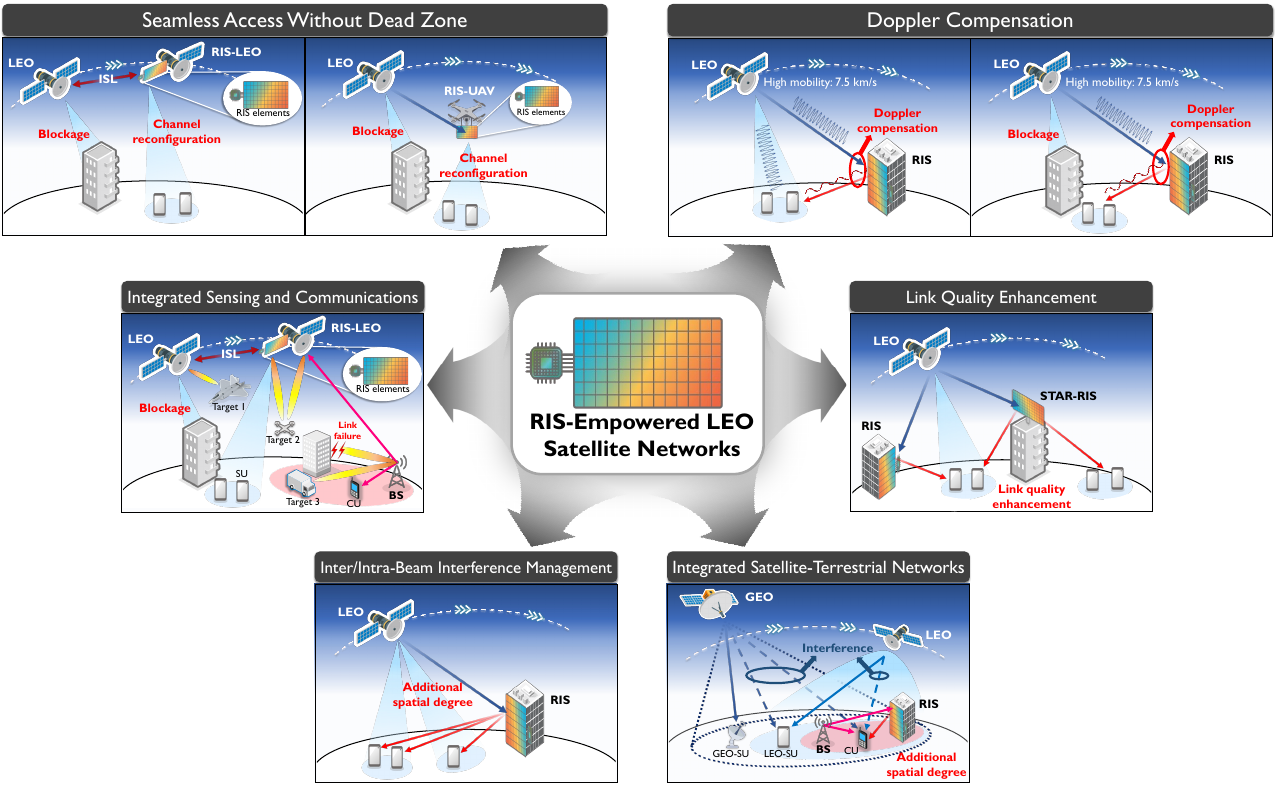}
    \end{center}
    \setlength{\belowcaptionskip}{-2pt}
    \caption{RIS-empowered scenarios in LEO satellite networks. SU: satellite user, and CU: cellular user.}
    \label{fig1}
\end{figure*}

\subsection{Potential Benefits}

\subsubsection{{Cost Effective and Power Efficient}}

RISs do not require any complicated signal processing operations, obviating the need for a radio-frequency (RF) chain and amplifier; thus, they consume much less power than other active devices, such as relays. RISs also offer low-cost fabrication because they are structurally composed of thin layers of lightweight material. However, to some extent, these beneficial features are negated by the components required for active RISs. {These RISs need additional amplifier units} 
increase the power consumption and cost as well as the thermal noise. Nevertheless, these additions are less significant than those of other conventional devices \cite{ZhangActiveRIS}. Additionally, RISs are cost-effective compared to phase arrays which are being used for satellites. This is because phase arrays utilize multiple RF chains and consist of high hardware complexity. Thus, thanks to enabling lighter payload and power efficiency, RISs can be adequately mounted on the main body and/or behind solar panels of LEOs with the limited cost and power budget of LEO \cite{Tekbiyik}. 



\subsubsection{{Less Transmission Delay}}

RISs have a full-duplex reflection structure in that a signal is simultaneously received and reflected in a one-time slot. Thus, the transmission delay is less than that of conventional relay-based units such as those operating in decode-and-forward and amplify-and-forward in two-time slots \cite{YeJia}. 


\subsubsection{{Noise-free Structure}}

Generally, since RISs are passive units without A/D or D/A converters or amplifiers, they do not add noise to the transmitted signal. {However, when the} 
{active RIS is considered, network designers should consider noise effects because of additional amplifier units.}  


\section{Promising Usage Scenarios and Applications for RIS-empowered LEO Satellite Networks}

To inspire future research in LEO-based 6G networks, we investigate promising usage scenarios and applications, as shown in Fig. \ref{fig1}. Specifically, RIS deployments can flexibly increase spatial dimension and adjust beam focusing for populated and unpopulated areas, respectively. Since RISs have in-band full-duplex capabilities ranging from sub-6 GHz to THz bands, the considered usage scenarios are applicable for VSAT (via Ku/Ka-bands) and D2C connectivity (via L/S-bands), and also valid for both downlink and uplink configurations.

\begin{figure*}[t!]
     \centering
     \begin{subfigure}[t]{0.65\textwidth}
         \centering
         \includegraphics[width=\textwidth]{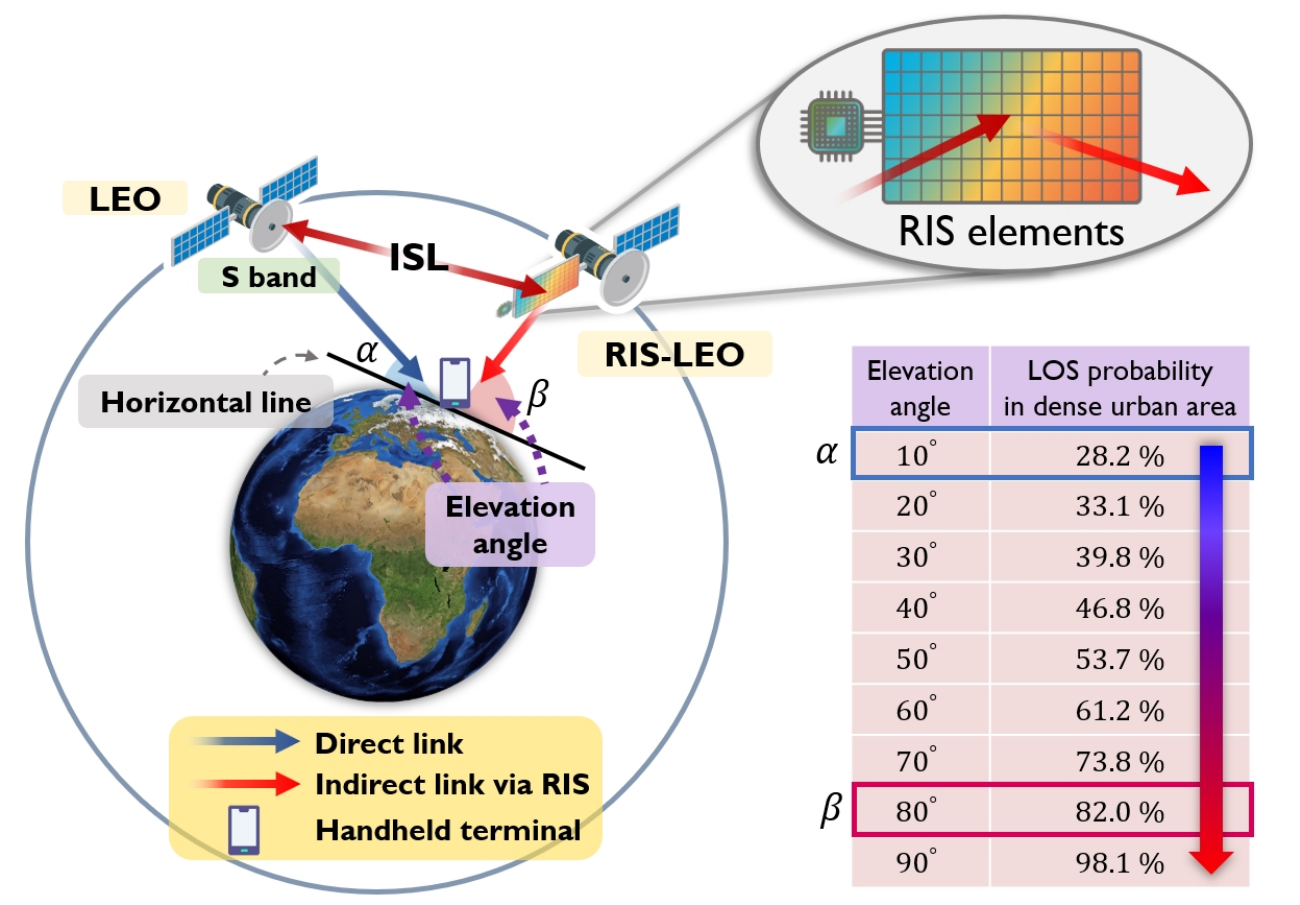}
         \caption{}
         \label{fig_Juhwan_a}
     \end{subfigure}
     \hfill
\begin{subfigure}[t]{0.65\textwidth}
    \centering
         \includegraphics[width=\textwidth]{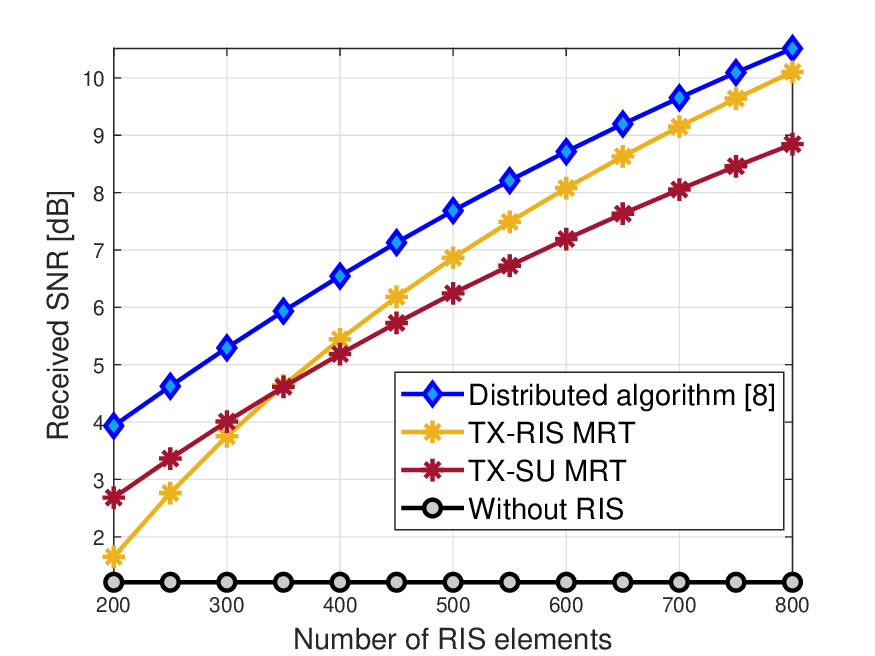}
         \caption{}
         \label{fig_Juhwan_b}
     \end{subfigure}
     \hfill
        \caption{RIS-assisted channel reconfiguration: (a) System model and elevation angle versus LOS probability and (b) Received SNR (dB) versus the number of RIS elements. 
        }
        \label{fig_Juhwan}
\end{figure*}

\subsection{{RIS-assisted Seamless Access Without Dead Zone} }

Due to obstructions on the ground, there may exist \textit{dead zone} areas that LEO satellites cannot cover. 
To achieve ubiquitous global connectivity without any dead zone, employing RIS on other LEO satellites or unmanned aerial vehicles (UAVs) can assist with information transmission, as depicted in Fig. \ref{fig1}. 
By controlling the reflection phases in RIS, an indirect link between the satellite and end-user with cascaded channels via RIS can be established to reconfigure wireless propagation and enhance the quality of service with coverage expansion.
%
%
In particular, for RIS-assisted UAVs, taking advantage of their mobility, higher LOS probability in channels between RIS and users can be provided to overcome the blockage problem.
%
%
%

Furthermore, in the areas with a low elevation angle, the channel between the LEO satellite and a satellite user (SU) has a low LOS probability.
In addition, shadowing and clutter loss are extremely high, leading to extremely poor data rates for the SU. 
To tackle this problem, \cite{Juhwan} proposed the use of RISs deployed on another LEO satellite with a higher elevation angle towards SU, thereby creating a favorable propagation environment, as shown in Fig.~\ref{fig_Juhwan}(a).
Therefore, we take account of the system model suggested in \cite{Juhwan} to be a case study. 
We consider an S-band spectrum in a downlink network, which provides moving beams on the satellite, and a handheld terminal as the SU in a dense urban area where the elevation angle is $10^\circ$.
For the RIS-to-SU link, lower path loss and improved link budget are guaranteed, along with the higher LOS channel probability. 
Since the ISL is assumed to be in free space, the LOS probability is almost $100\%$. 
Considering the long propagation distance, severe path loss, and hardness of real-time channel state information (CSI) exchange via ISL, a distributed algorithm \cite{Juhwan} based on alternating optimization is utilized. Herein, the transmit and RIS beamformings are alternatively optimized to maximize the signal-to-noise ratio (SNR) at the SU in an iterative manner, providing closed-form solutions with low computational complexity in each iteration.
To demonstrate the superiority of the distributed algorithm, we illustrate performance results in Fig.~\ref{fig_Juhwan}(b) with comparisons to the benchmark schemes. 
``{\sf TX-RIS MRT}'' and ``{\sf TX-SU MRT}'' represent the schemes where the transmit beamforming is set to maximum ratio transmission (MRT) based on the channel of LEO-to-RIS link, and that of LEO-to-SU link, respectively. Here, the RIS phase shifts are optimized using the distributed algorithm. 
``{\sf Without RIS}'' denotes the scheme with only the direct link from the transmitter to the SU without indirect links via RIS.
We can see that the distributed algorithm outperforms the benchmark schemes; when the number of RIS elements is 800, the received SNR via the distributed algorithm is about 9 dB higher than that of the case without RIS. 
This is due to the establishment of cascaded channels via RIS, which guarantees a much higher LOS probability. Furthermore, the distributed algorithm takes account of the phase shifts of RIS and channel effects of RIS-SU links to find the transmit beamforming, while not considered in the other schemes.

\subsection{RIS-assisted Doppler Compensation}

Given the exploitation of higher frequencies and the large spectrum in LEO satellite networks, the Doppler shift has an immense effect.
To overcome this problem, various approaches have been proposed in  pre-/post-compensation manners, namely, increasing the sub-carrier spacing of an orthogonal frequency division multiplexing (OFDM) signal, using a phase-locked loop to detect both the carrier frequency and phase bias, estimating the Doppler shift by pilot signals and/or AI-based techniques, using orthogonal time-frequency space (OTFS) modulation, and so on. 
Although the approaches mentioned above partially compensate for the Doppler shift, they introduce additional problems, i.e., cyclic prefix overhead, low spectrum efficiency, high computational complexity, and increased latency. 




Thus, it is reasonable to utilize RISs owing to their remarkable benefits explained in Section III-B. By doing so, the Doppler shift can be compensated for at a high level, and the hardware complexity of the end-user (as well as the satellite) can be reduced, while the signal quality can be increased \cite{BasarRISDoppler}. In this context, the compensation can be either direct or indirect. Direct-compensation involves determining the optimal RIS configuration by appropriately adjusting the phase shifts of the RIS elements in the presence of the Doppler effect. To this end, channel phases should be effectively estimated, and the Doppler shift information is required. This approach not only compensates for the Doppler shift, but the received SNR gain also significantly increases \cite{MattBho}. Indirect-compensation entails identifying appropriate phase shifts by focusing on compensating for only the channel phases, regardless of the Doppler effect. In this approach, the RIS does not need the information of the Doppler shift, because the focus is on maximizing the received signal strength via compensating for the channel phases. In both cases, RISs can be active, passive, and/or dynamic. Subsequently, two main types of usage scenarios arise, as shown in Fig. \ref{fig1}. {For the scenario shown on the left, only the Doppler shift in satellite-to-RIS link can be compensated because the RIS node cannot control the signal in satellite-to-user link.} Additionally, for the scenario shown on the right, it is possible to significantly compensate for the Doppler shift using direct/indirect compensation methods \cite{BasarRISDoppler}.

\begin{figure}[t!]
    \begin{center}
        \includegraphics[width=0.8\textwidth]{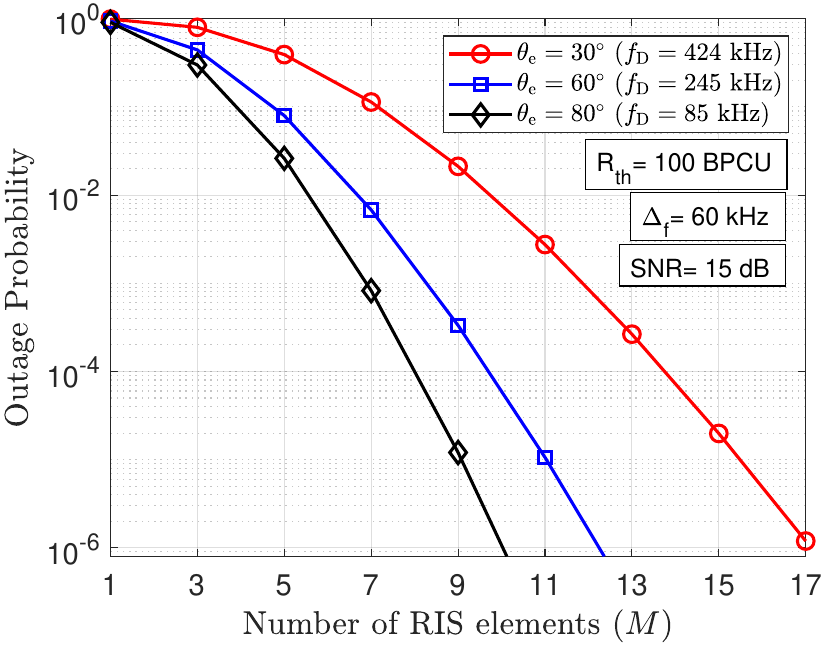}
    \end{center}
    \setlength{\belowcaptionskip}{-2pt}
    \caption{OP performance of RIS-aided LEO network under frequent heavy shadowing environment.}
    \label{fig1_1}
\end{figure}

To demonstrate performance gains offered by RIS, we evaluate the outage probability (OP) of the scenario shown on the right in Fig. \ref{fig1}. We consider an OFDM-based LEO satellite downlink network utilizing Ka-band. Each narrowband sub-channel in satellite links (LEO-to-RIS) is distributed as independent and identically distributed shadowed-Rician fading \cite{AbdiA}. Terrestrial links (RIS-to-user) are assumed to be distributed as Rayleigh fading. Other parameters are carrier frequency ($20$ GHz), bandwidth ($245.76$ MHz), LEO altitude ($\num{1000}$ km), the antenna gain of LEO ($24$ dB), and noise temperature at user $500^{\circ}$ K. The number of sub-carriers in the OFDM scheme $\num{4096}$, and $\Delta_\text{f}$ denotes sub-carrier spacing. Further, $f_\text{D}$, $R_\text{th}$, and $\theta_\text{e}$ represent the maximum Doppler frequency, bits per channel in use (BPCU), and elevation angle, respectively. 
As observed in Fig. \ref{fig1_1}, the OP performance severely degrades with decreasing elevation angle (increasing Doppler frequency) for a fixed number of RIS elements. 
The performance can be significantly improved, and an OP value of $10^{-6}$ can be achieved with fewer RIS elements in the case of $\theta_\mathrm{e}=30^{\circ}$. The figure shows that RISs play a promising role in improving performance through overcoming the Doppler effect.


\subsection{RIS-assisted Link Quality Enhancement}

In a satellite-terrestrial link, channel quality can be poor due to masking effects caused by the atmosphere, path loss, and obstacles. To overcome this, the concept of exploiting terrestrial relays in satellite communication has been widely considered. However, this increases cost, power consumption, and latency because of its inherent hardware structure. 

In contrast, RISs are more capable of improving link quality than relays.
Furthermore, RISs do not require spectral resources to be divided into orthogonal resources; thus, they have full-duplex structures. 
As seen in Fig. \ref{fig1}, two approaches can be used to improve link quality using RISs. The first one is to deploy conventional RISs on ground buildings. Thus, signal quality received by SUs can be significantly improved in an urban area. In this case, only SUs located within the half-space reflection area can enjoy the assistance of RISs. In the case of a huge blockage, the second approach is to utilize STAR-RISs. Thereby, each SU located within $360^{\circ}$ coverage of STAR-RIS can enjoy link improvements. Consequently, with/without direct LOS, both approaches significantly enhance link quality, together with additional benefits in LEO networks. 


\subsection{RIS-aided Inter-/Intra-beam Interference Management} 
To increase the spectrum efficiency (SE) according to high-throughput demands for given limited spectrum resources, LEO satellites should aggressively reuse frequency bands; however, this results in inter-beam interference for beam edge users and intra-beam interference for users within the beam.
Various MA techniques have been leveraged to mitigate such interference by creating beams at different spatial directions.
To appropriately manage interference using such MA techniques, a sufficient spatial resource is required at LEO satellites. However, in satellite systems, the number of users within the coverage area is considerably large since satellites serve wide geographic regions. Along with this, {DVB-S2X} based multibeam multicast transmission is adopted to overcome severe fading loss. Therefore, satellite networks are usually user-overloaded in that the number of users within the wide coverage exceeds that of transmission antennas.
To address these limitations, deploying RISs within the coverage area, as shown in Fig. \ref{fig1}, can be one of the solutions. 
In this setup, the RIS can reflect the signal toward the SUs by appropriately adjusting the phases of the RIS elements to nullify the interference signals and boost up the desired signal \cite{YeJia}.
In light of this, hybrid beamforming that simultaneously optimizes the phase of RIS elements and the precoder at the LEO satellite can be performed. This in turn provides additional \textit{spatial dimensions} to increase the SE effectively.

{However, this gives rise to possible challenges that must be addressed.}
Since the channel quality of the RIS-SUs depends on the distance between the RIS and SUs, multiple RISs need to be deployed within the coverage to enhance the individual SE uniformly. This may cause additional inter-RIS interference because the optimized phase shifts between any arbitrary SU and the RIS could interfere with other SUs.
To mitigate these impairments, cooperatively integrating multiple RISs may be a solution; however, the exchange of control information among multiple RISs introduces processing overhead.

\subsection{{RIS-aided Integrated Satellite-Terrestrial Network} }

Integrating satellite and terrestrial networks (ISTNs) have received considerable attention from academia and industry to increase coverage, reliability, and spectrum efficiency. Two prominent key designs are available for the ISTN concept, namely, hybrid-ISTN (H-ISTN) and spectrum-sharing ISTN (SS-ISTN), as shown in Fig. \ref{fig1}. In the H-ISTN design, terrestrial infrastructures are utilized as relay nodes to overcome signal degradation resulting from heavy masking effects. However, as explained in Section III-B, RISs are competitively preferable to terrestrial relays because of their promising advantages. On the other hand, the SS-ISTN design enables satellites and terrestrial infrastructures to jointly exploit the same spectrum resources in a cognitive-radio manner (or not) \cite{YeJia}. 
However, LEO-SUs and CUs suffer from co-channel interference (CCI) coming from GEO and LEO, respectively, in a coexistence network of LEO and GEO. CUs also experience interference from GEO. Furthermore, in a large-scale LEO constellation, many deployed LEOs inflict excessive CCI on GEO-SUs. Fortunately, it is possible to mitigate the interference signals by deploying multiple RISs dedicated to GEO-SUs, LEO-SUs, and CUs. Nevertheless, this remains challenging because appropriate phase adjustment must be guaranteed.

\subsection{RIS-aided Integrated Sensing and Communications}

With the ever-increasing number of Internet-connected devices, ISAC has been regarded as a potential solution to utilize the predetermined limited radio spectrum efficiently. In ISAC, sensing and communication are jointly maintained via a shared spectrum; thus, it can increase the spectral/energy efficiency, lower the hardware cost, and decongest the crowded spectrum not only in terrestrial but also in LEO satellite networks.

Besides, an RIS also can sense a target along with enhancing the ability of ISAC. For instance, as depicted in Fig. \ref{fig1}, in the absence of LOS, in which the sensing is generally quite challenging, the BS requests the assistance of the RIS-LEO to sense Target-2. Moreover, if the RIS-LEO cannot sense Target-1 for the BS, the LEO (without RIS) can be in charge and inform the RIS-LEO via ISL. When LOS exists, the RIS (whether mounted on buildings or LEOs) can enhance sensing and localization accuracy as well as increase the resolution. With the aid of terrestrial RIS and/or RIS-LEO, both the sensing accuracy can be increased, and communication service continuity can be ensured in ISAC-based LEO networks and/or ISTN. However, with/without the aid of RISs, appropriate resource and interference management are crucial for sensing and communication. Although recent studies have attempted to address this issue via beamforming, waveform designs, and MA techniques, it remains an open problem.

\section{Future Directions and  Challenges} 
\label{sec:4}

All deployment configurations and use cases of RIS-empowered LEO satellites are foreseen to play a vital role in global connectivity under 5G-advanced and 6G standards. In this context, the following subsections present potential future research directions and possible challenges.

\subsection{Potential Future Directions}

\subsubsection{AI-enabled RIS-aided LEO Satellites}

Since beamforming designs for RIS units require high computational complexity; thus, energy- and spectral-efficient approaches that implement AI-assisted RIS implementation can be expected to play a vital role. In this regard, \cite{9593256} considered a neural network-based solution for a spectral-efficient RIS design. Furthermore, \cite{10032271} deals with a deep deterministic policy gradient approach consisting of several deep neural networks to develop an intelligent and energy-efficient resource allocation scheme.

\begin{figure}[t!]
         \centering        \includegraphics[width=0.75\textwidth, trim=10 10 0 0]{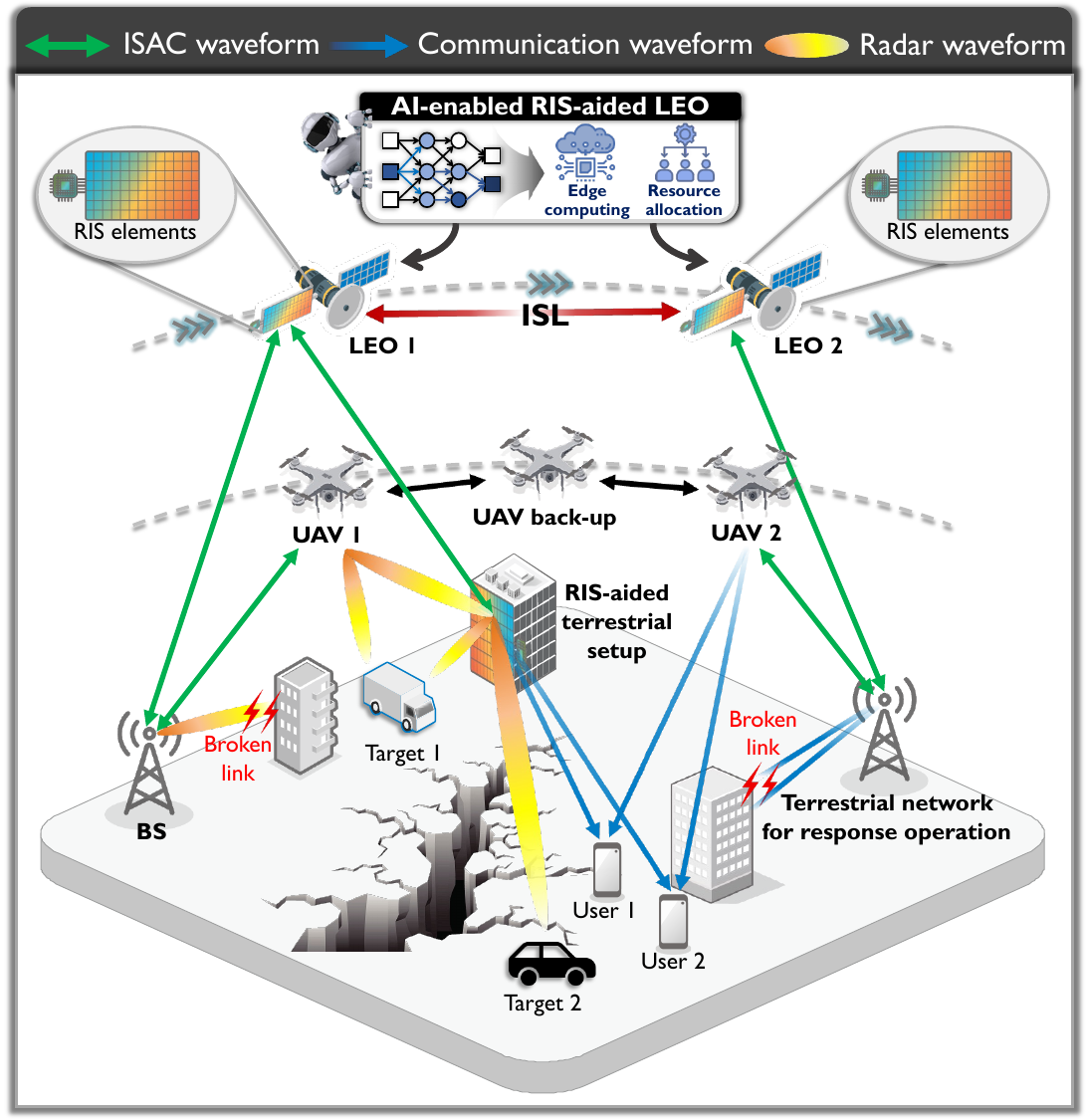}
         \caption{AI-enabled RIS-aided LEO satellites with ISAC functionalities for public safety operations.}
         \label{fig_RIS_LEO}
\end{figure}

In Fig. \ref{fig_RIS_LEO}, we present a futuristic illustration of AI-enabled RIS-empowered LEO satellites that can be employed with ISAC functionalities for public safety operations. Edge computing architectures consisting of the terminal–satellite–cloud, where tasks can be processed at three planes and inter-satellites can cooperate to achieve on-board load balancing, can be implemented for intelligent resource allocation using AI methods as shown in Fig. \ref{fig_RIS_LEO}. Also, the use of AI methods for channel training becomes quite important since channel training is required for high-resolution sensing, tracking, and 3D localization. In Fig. \ref{fig_RIS_LEO}, LEO satellites and UAVs act as relays to sense targets and communicate with users in a disaster-affected region. The satellite LEO-1 transmits a dual-function ISAC signal that is reflected toward targets and users through an RIS-aided setup on the terrestrial network that is not disaster-affected. The RIS embedded in LEO-1 helps to transfer information and sensing data with the LEO-2 satellite via an ISL. The LEO-2 satellite, with its embedded RIS setup, can further transmit the ISAC signal to a terrestrial network-based control unit responsible for post-disaster response operations. To enable AI-based LEO satellite applications empowered by the RIS, advanced and efficient learning approaches with low training overhead are required, which have yet to be explored in the literature.

\subsubsection{Holographic Multiple-Input Multiple-Output (MIMO) Surfaces at THz for LEO Satellites}

In addition to RISs, reconfigurable holographic MIMO surfaces (HMIMO), where the transceiver leverages the hologram principle for efficient communication and networking applications, have advanced. 

The radio meta-surface of an HMIMO-aided transceiver does not require an additional control unit to phase-shift the holographic transmit signal \cite{9136592}. 
Both the RIS and HMIMO configurations can be applied across the radio spectrum, from sub-6 GHz through THz frequencies with massive antenna connectivity. RIS-aided systems have been exploited for LEO satellite networks at THz bandwidth \cite{YeJia}. However, more advanced cases of HMIMO have yet to be developed for LEO satellite networks.  
Under the realm of the massive IoT use cases of 5G-advanced, it is vital to design green HMIMO designs with optimization and low-complexity AI tools. 



\subsection{Practical Challenges}

\subsubsection{{Channel Estimation}}

Because RISs have passive structures, obtaining accurate CSI of transmitter-RIS and RIS-user links is impossible. Two approaches have been widely considered in the literature. One is deploying receive RF chains with channel sensing ability and low power consumption. With this approach, CSIs in both links can be estimated individually. The second is to estimate end-to-end effective CSI to adjust the phases of the RIS unit. 
However, neither approach guarantees the accuracy of the CSI, which is crucial for obtaining the full benefits of RIS in LEO networks. To obtain more accurate CSIs in RIS-enabled communications, leveraging deep learning to estimate end-to-end CSIs, as well as transmitter-RIS and RIS-user CSIs, has drawn great attention.


\subsubsection{{Phase Uncertainty}}

Phase uncertainty refers to imperfections in phase estimation and/or phase quantization. For instance, if the phase of an element in the RIS cannot be adjusted to maximize the effective end-to-end SNR, the remaining channel phase will be considered phase noise at user-ends. Furthermore, because the RIS consists of many elements, phase noise aggregates at the user-ends, which limits the benefits of the spatial DOF gained by the RIS. To prevent such issues, shifting the RIS phase elements to be averagely robust to phase errors by characterizing statistical information of phase error can be a potential method.


\subsubsection{{Electromagnetic Interference Radiation}}

RISs reflect not only the desired signals but also undesired environmental signals; thus, this causes electromagnetic interference (EMI) effects at the user-ends. Khaleel \textit{et al.} \cite{KhaleelEMI} drew the attention of researchers to how EMIs can be mitigated. They show that the effect of EMIs can be canceled at the cost of sacrificing some degree of passive beamforming gain; however, the performance can be significantly improved in terms of the average received signal-to-interference-to-noise ratio. 

\subsubsection{Mobility Management}

High mobility severely affects the performance of RIS-aided networks from different perspectives, particularly for LEO-based networks. In the case of high mobility, variations in channels change rapidly; this causes the estimated channels to be outdated (in other words, outdated CSI), as well as beam misalignments. To overcome this, frequent pilot transmission may be required at the cost of sacrificing the transmission rate. Most recently, machine learning-based approaches have received much interest in RIS-aided LEO networks for estimating more precise CSIs with reduced pilot overhead and computational complexity.


\subsubsection{Unified Multi-layer Architecture}

To provide seamless coverage, large RIS-LEO constellations must be combined with GEO and/or UAV using coverage-aware control signaling. It is essential to implement a multi-layered architecture for service continuity and reinforced reliability/availability
in regions that have made it difficult for GEO satellites to provide communication services so far. In addition, an integrated network protocol that includes multi-layer routing and resource allocation should be introduced by using the flying RIS.

\section{Conclusion}

This article has provided a holistic overview of LEO satellite communication and RIS technology to understand why we must utilize RIS in LEO satellite networks towards 6G. We next focused on promising usage cases and challenges, together with numerical results demonstrating the benefits of the RIS technology in LEO satellite communications. Finally, we have discussed the potential future directions and challenges to foster research for forthcoming NTN-based 6G networks.



\bibliographystyle{IEEEtran}

\bibliography{IEEEabrv,BibRef}

 \begin{IEEEbiographynophoto}
{Mesut Toka}(Member, IEEE) received his Ph.D. degree in Electronics Engineering at Gebze Technical University (GTU), Kocaeli, T\"{u}rkiye, in 2021. He was a Research and Teaching Assistant at GTU during his Ph.D. period. He is currently working as a Postdoctoral Researcher in Electrical and Computer Engineering at Ajou University, Suwon, South Korea. 
 \end{IEEEbiographynophoto}\vspace{-7mm}
 \begin{IEEEbiographynophoto}
 {Byungju~Lee} (Member, IEEE)
received a Ph.D. degree from Korea University, South Korea, in 2014, and then was a Postdoctoral Scholar at Purdue University, USA. He is currently an Assistant Professor at Incheon National University, South Korea. Dr. Lee was awarded the Fred W. Ellersick Prize.
 \end{IEEEbiographynophoto}\vspace{-7mm}
   \begin{IEEEbiographynophoto}
  {Jaehyup~Seong} (Graduate Student  Member, IEEE) is currently pursuing a Ph.D. degree in the School of Electrical Engineering at Korea  University, South Korea.
 \end{IEEEbiographynophoto}\vspace{-7mm}
\begin{IEEEbiographynophoto}{Aryan Kaushik} (Member, IEEE) has been with the University of Sussex, UK, since 2021. Previously, he has been with UCL, University of Edinburgh, and HKUST. He is the Editor of two books, IEEE CTN, IEEE OJCOMS (Best Editor Award 2023), IEEE COMML (Exemplary Editor 2023), and several IEEE special issues.
\end{IEEEbiographynophoto}\vspace{-7mm}
\begin{IEEEbiographynophoto}
  {Juhwan~Lee} (Graduate Student  Member, IEEE) is a Ph.D. candidate in the Department of Electrical and Computer Engineering at Seoul National University, South Korea.
 \end{IEEEbiographynophoto}\vspace{-7mm}
  \begin{IEEEbiographynophoto}{Jungwoo Lee} (Senior Member, IEEE)
 received M.S.E. and Ph.D. from Princeton University. He is now a professor at ECE, Seoul National University. He received the Qualcomm Dr. Irwin Jacobs award in 2014. He is the co-recipient of the 2020 IEEE Fred W. Ellersick prize. He is a general co-chair for ISIT 2028.
  \end{IEEEbiographynophoto}\vspace{-7mm}
  \begin{IEEEbiographynophoto}{Namyoon~Lee} (Senior Member, IEEE)
received a Ph.D. degree from The University of Texas at Austin, USA, in 2014, and worked as a Research Scientist at Intel Labs, USA. He is currently an Associate Professor at the School of Electrical and Electronic Engineering, Korea University, South Korea.  \end{IEEEbiographynophoto}\vspace{-7mm}
   \begin{IEEEbiographynophoto}{Wonjae Shin} (Senior Member, IEEE)
 received a Ph.D. degree from Seoul National University, South Korea 2017, and then was a postdoctoral research fellow at Princeton University, USA. He is currently an Associate Professor at Korea University, South Korea. Dr. Shin was awarded the Fred W. Ellersick Prize and the ComSoc Asia-Pacific Young Researcher Award.
 \end{IEEEbiographynophoto}\vspace{-7mm}
\begin{IEEEbiographynophoto}{H. Vincent Poor} (Life Fellow, IEEE) received the Ph.D. degree in EECS from Princeton University, Princeton, NJ, USA, in 1977. Since 1990, he has been on the faculty at Princeton, where he is the Michael Henry Strater University Professor of Electrical Engineering. From 2006 until 2016, he served as a Dean of Princeton's School of Engineering and Applied Science. 
  \end{IEEEbiographynophoto}

\end{document}